\documentclass[aps,prd,twocolumn,superscriptaddress,showpacs,floatfix]{revtex4}
\usepackage[dvips]{graphicx}
\usepackage{natbib}

\newcommand{\mnras}{Mon. Not. R. Astron. Soc.}

\newcommand{\apjl}{Astrophys. J. Lett.}

\newcommand{\rhos}{\rho_\mathrm{s}}
\newcommand{\rs}{r_\mathrm{s}}
\newcommand{\vir}{\mathrm{vir}}
\newcommand{\Mvir}{M_{\mathrm{vir}}}

\newcommand{\Rvir}{R_{\mathrm{vir}}}
\newcommand{\Cvir}{C_{\mathrm{vir}}}  

\newcommand{\mDM}{m_{\mathrm{DM}}}  
\newcommand{\mboson}{m_{V}}

\newcommand{\vchar}{v_{\mathrm{char}}}
\newcommand{\vesc}{v_{\mathrm{esc}}}
\newcommand{\vrel}{v_{\mathrm{rel}}}
\newcommand{\sigmav}{\sigma_{\mathrm{v}}}
\newcommand{\sigmavz}{(\sigma v)_{0}}

\newcommand{\avsigmav}{\langle\sigma v \rangle}

\newcommand{\dd}{\mathrm{d}}
\newcommand{\kpc}{\mathrm{kpc}}

\newcommand{\Msun}{M_{\odot}}

\begin{document}

\title{
Dark matter annihilation rates with velocity-dependent annihilation cross sections
}
\author{Brant E. Robertson}
\affiliation{
Kavli Institute for Cosmological Physics, and Department of Astronomy and
Astrophysics, University of Chicago, 933 East 56th Street, Chicago, IL 60637, USA
}
\affiliation{
Enrico Fermi Institute, 5640 South Ellis Avenue, Chicago, IL 60637, USA
}
\author{Andrew R. Zentner}
\affiliation{
Department of Physics and Astronomy, University of Pittsburgh, Pittsburgh, PA 15260, USA
}

\date{\today}

\begin{abstract}

The detection of byproducts from particle annihilations 
in galactic halos would provide important information about
the nature of the dark matter.  Observational evidence for a 
local excess of high-energy positrons has motivated recent models 
with an additional interaction between dark matter particles that 
can result in a Sommerfeld enhancement to the cross section.
In such models, the cross section for annihilation becomes velocity-dependent and 
may enhance the dark matter annihilation rate in the solar neighborhood 
relative to the rate in the early universe sufficiently to 
source observed fluxes of high-energy positrons.  We demonstrate that, for
particle interaction cross-sections that increase with decreasing velocity,
the kinematical structures of dark matter halos with interior 
density profiles shallower than isothermal, 
such as Navarro-Frenk-White or Einasto halos, may induce a 
further enhancement owing to the position-dependent velocity distribution.
We provide specific examples for the increase in the annihilation rate with
a cross-section enhanced by the Sommerfeld effect.  In dark matter halos like 
that of the Milky Way and Local Group dwarf galaxies, the effective cross section at 
the halo center can be significantly larger than its local value.  
The additional enhancement owing to halo kinematics depends upon the 
parameters of any model, but is a prediction of certain models aimed at 
explaining measured positron fluxes and can exceed an order of magnitude.

\end{abstract}

\pacs{98.35.+d, 98.62.Gq, 98.35.Gi, 14.80.Ly}

\maketitle

\section{Introduction}
\label{section:introduction}
%
%

Astrophysical probes are important elements of the effort 
to identify the cosmological dark matter (DM).
Detecting $\gamma$-rays from DM annihilation is
a prime scientific motivation for existing and forthcoming
$\gamma$-ray instruments including the space-based Fermi 
Gamma-ray Space Telescope (FGST) \cite{ritz2007a}, and atmospheric detectors such as 
the High Energy Stereoscopic System (HESS) \cite{hofmann2003a}, 
the Very Energetic Radiation Imaging Telescope Array System (VERITAS) \cite{weekes2005a}, 
the Major Atmospheric Gamma-ray Imaging Cerenkov telescope 
(MAGIC \& MAGIC II) \cite{bastieri2008a}, the 
Collaboration of Australia and Nippon for 
Gamma-ray Observatory in the Outback (CANGAROO) \cite{kubo2004a}, 
the High-Altitude Water Cerenkov experiment (HAWC) \cite{sinnis2005a}, 
and the Whipple 10m \cite{aharonian2006a,aharonian2008a,ritz2007a,wood2008a}.
The unexpected positron excess above $~10$~GeV detected by 
the Payload for Anti-matter Exploration and Light-Nuclei Astrophysics (PAMELA) 
satellite \cite{adriani2008a} and the cosmic ray 
electron/positron excess at $\sim 300-800$~GeV detected by the 
Advanced Thin Ionized Calorimeter (ATIC) instrument 
\cite{chang2008a} have been interpreted as products of 
DM annihilation in the halo of the Milky Way and 
have driven significant interest in leptophillic particle DM 
models with suppressed hadronic production channels 
(e.g., Refs.~\cite{cirelli2008a,arkani-hamed2008a}).

The DM annihilation interpretation of the PAMELA and ATIC 
data faces a challenge in that the required cross section  
is roughly an order of magnitude or more larger than 
the value $\avsigmav \sim 3\times 10^{26}~\mathrm{cm}^{3}~\mathrm{s}^{-1}$ 
necessary for a thermal relic to have the observed contemporary 
DM density \cite{cirelli2008a,cholis08,donato08,ishiwata08}.  
A class of proposals that gives rise to large effective cross sections 
in galactic halos introduces an attractive force among dark 
matter particles mediated by a relatively light boson.  The annihilation 
cross section can be large courtesy of a non-perturbative correction 
called the ``Sommerfeld enhancement'' 
\cite{sommerfeld1931a,bergstrom89,hisano_etal04,hisano_etal05,profumo05,arkani-hamed2008a,lattanzi2008a,march-russell2008a,march-russell2008b,chen2009a}.  
The new force effectively focuses incident plane-wave wavefunctions 
and can result in a significant increase in
the effective annihilation cross section.
The sizes of these ``Sommerfeld-enhanced'' cross sections are very sensitive 
to the parameters of the model and are generally monotonically decreasing 
functions of relative encounter speed \cite{arkani-hamed2008a,lattanzi2008a}.

While the antiparticle excesses may have a more
pedestrian astrophysical origin \cite{busching_etal08,profumo2008a},
Sommerfeld-enhanced DM annihilation yields interesting phenomenology.  
Annihilation rates scale as $\Gamma \propto \avsigmav \rho^2$ where 
$\avsigmav = \int f(v) \sigma(v)v\ \dd v$ is the cross section times velocity 
averaged over the distribution of particle velocities $f(v)$, and $\rho$ is the 
local density.  In canonical DM particle scenarios, 
$\sigma(v)v$ is constant at $v/c \ll 1$ and the integral is trivial.  
Previous estimates of the annihilation rates in Sommerfeld-enhanced 
models adopted a single characteristic velocity $\vchar$ to describe entire halos, 
approximating $\int f(v) \sigma(v)v\ \dd v \approx \sigma(\vchar)\vchar$.  
This approach may sensibly address the local PAMELA/ATIC data but 
neglects other observational consequences of such models.  Moreover, while
the constant halo velocity approximation is valid in the case of isothermal 
halos where density depends upon radius as $\rho \propto r^{-2}$, the structures of 
dark matter halos are generally not isothermal (and not Maxwellian) 
and the velocity distribution will depend upon position.  

The spatially-dependent velocity scale in DM halos is intuitive.
Cosmological simulations suggest that DM halos have density profiles 
that scale as $\rho(r) \sim r^{-1}$ in their interiors 
(e.g., Ref.~\cite{dubinski1991a,navarro1996a}).  
The characteristic speeds of particles then scale 
as $\vchar \sim \sqrt{G M(<r)/r} \propto \sqrt{r}$, where 
$M(<r)$ is the mass contained within a sphere of radius $r$.  
Characteristic relative velocities decrease with radial position.  As 
a consequence, the velocity dependence of the cross section can result in a position 
dependence in the annihilation rate beyond the 
standard $\Gamma \propto \rho^2(r)$ proportionality.  
We use the Sommerfeld enhancement to provide a popular example of 
a velocity-dependent cross section, but we emphasize that the 
simple point of this paper pertains to any model in which the 
annihilation cross-section is velocity dependent.  Models in which 
s-wave annihilation is forbidden are another class of scenarios 
where annihilation cross sections can be velocity dependent, and 
such scenarios have been studied within the context of dark matter 
(a recent example in a similar context is Ref.~\cite{boehm_etal04}).

To calculate annihilation rates in cases of velocity-dependent 
interaction cross sections, an average over the radially-dependent velocity 
distribution should be performed.  We perform this calculation and provide 
examples of the modified DM annihilation rates below.  In particular 
we show that for some choices of parameters, the radial dependence 
of the velocity distribution can lead to important modifications 
to the radial dependence of the annihilation rate.

\section{Halo Density and Velocity Structure}
\label{section:halo_profiles}

\citet{navarro1996a} (NFW) found that the density profiles of
DM halos in cosmological simulations scaled
as $\rho(r)\propto r^{-1}$ at small radii and transitioned to
$\rho(r)\propto r^{-3}$ in the halo exterior.  We take the NFW 
density profile, 
\begin{equation}
\label{equation:nfw}
\rho(x)=\frac{4\rhos}{x(1+x)^{2}},
\end{equation}
\noindent
where $x\equiv r/\rs$ as one of our example halo models.  
The NFW profile has a scale density $\rho(x=1)=\rhos$.

\begin{figure*}
\begin{center}
\includegraphics[height=8cm]{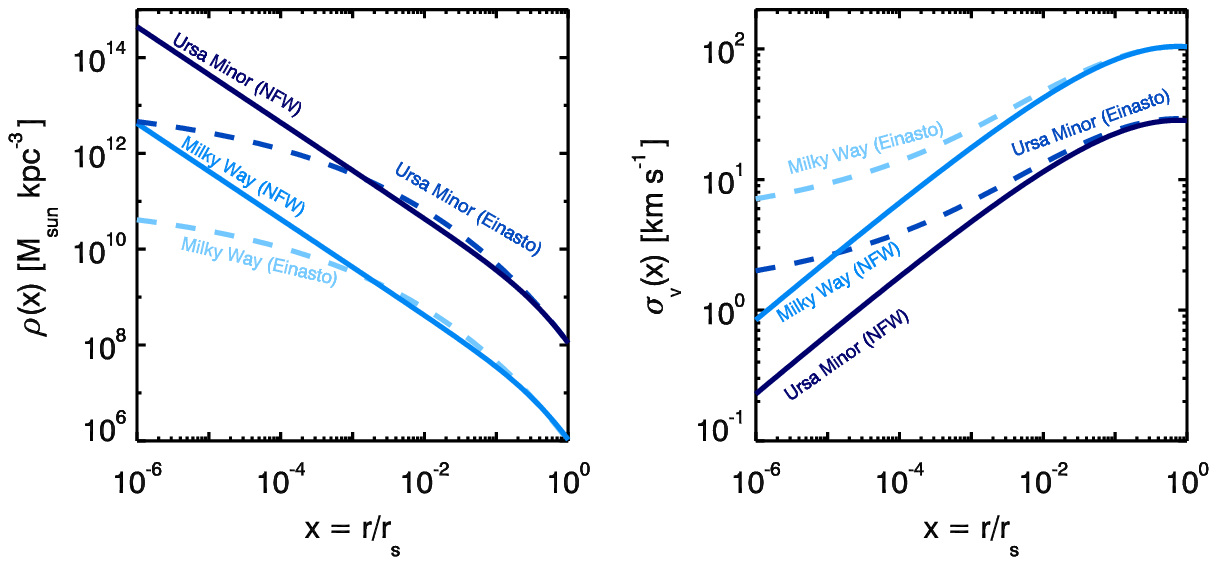}
\caption{\label{fig:halo_info}
Halo density (left panel) and velocity dispersion (right panel) profiles as a function 
of radius in units of halo scale radii, $x=r/\rs$.  We show 
NFW ({\em solid lines}) and Einasto ({\em dashed lines}) models of a Milky Way-like halo with 
parameters $M_{\vir} \approx 10^{12}h^{-1} \Msun$, and $\rs \approx \mathrm{kpc}$ \cite{klypin2002a} 
as well as the Local Group dwarf galaxy Ursa Minor with 
$\rhos \approx 1.1\times10^{8} \Msun~\mathrm{kpc}^{-3}$, and $\rs = 0.6~\mathrm{kpc}$ 
\cite{strigari2008a}.
}
\end{center}
\end{figure*}

More recent numerical studies have demonstrated that the innermost density
profiles of DM halos follow a radially-dependent power-law slope that
may soften relative to the NFW $\gamma=1$ value.
\citet{navarro2004a} find that the function
\begin{equation}
\label{equation:einasto}
\rho(x) = \rho_{s}\exp\left[-\frac{2}{\eta}\left(x^{\eta}-1\right)\right]
\end{equation}
\noindent
with $\eta \sim 0.17$ represents well the DM halos formed in 
their simulations.  This function
has a form similar to the distribution suggested 
by \citet{einasto1969a}, and we refer to Equation~\ref{equation:einasto} 
as the Einasto profile.  This density profile has a local 
power-law index $\dd \ln \rho/\dd \ln r\vert_{x=1} = 2$ 
and density $\rho(x=1)=\rhos$.

The left panel of Figure~\ref{fig:halo_info} shows the
density profile for two example galactic DM halos.
Constraints on the rotation curve of the Milky Way (MW) 
suggest that its DM halo has a virial mass of $\Mvir \approx 10^{12} \Msun$, 
halo concentration $\Cvir \equiv \Rvir/\rs \sim 12$,
and scale length $\rs \sim22$~kpc \cite{klypin2002a}.  
These parameters imply a DM density at the MW scale radius of
$\rhos \sim 10^{6}\Msun~\kpc^{-3}$.  
In addition to the 
MW, we consider annihilation in the halo of the Local Group satellite 
galaxy Ursa Minor because it subtends a large angle on the sky 
and should be among the most luminous dwarf galaxies in DM annihilation 
products \cite{strigari2008a}.  
Stellar kinematics 
constrain the Ursa Minor halo to have $\rs \approx 0.6~\kpc$ 
and $\rhos \approx 1.1 \times 10^{8} \Msun~\kpc^{-3}$ 
\cite{strigari2008a}.

For a spherically-symmetric system with an isotropic 
distribution of velocities, the equilibrium one-dimensional velocity 
dispersion $\sigmav$ can be computed from the Jeans equation 
\cite{binney1987a}, 
\begin{equation}
\sigmav^2(r) = \frac{1}{\rho} \int_{\infty}^{r}\ \rho \frac{\dd \Phi}{\dd r}\ \dd r,
\label{equation:jeans_sigmav}
\end{equation}
\noindent
where $\Phi$ is the total gravitational potential supplied by all
components of the galaxy, including structures
other than that for which the 
dispersion is being computed.  
The right panel of Figure~\ref{fig:halo_info} shows the velocity
dispersion $\sigmav(x)$ for models of the MW and Ursa Minor
DM halos modeled with NFW and Einasto density profiles.  
The NFW halo models reach a density $\sim 100 \times$ larger than the Einasto
profile models at $x \sim 10^{-6}$, and have significantly lower inner 
velocity dispersions.  Halos with structures similar to the Ursa Minor
satellite have halo velocity dispersions $\sim 5$ times smaller than the 
MW.  In what follows, we adopt these values as fiducial parameters for the 
halos of the MW and Ursa Minor.

%
%
\begin{figure*}
\begin{center}
\includegraphics[height=15cm]{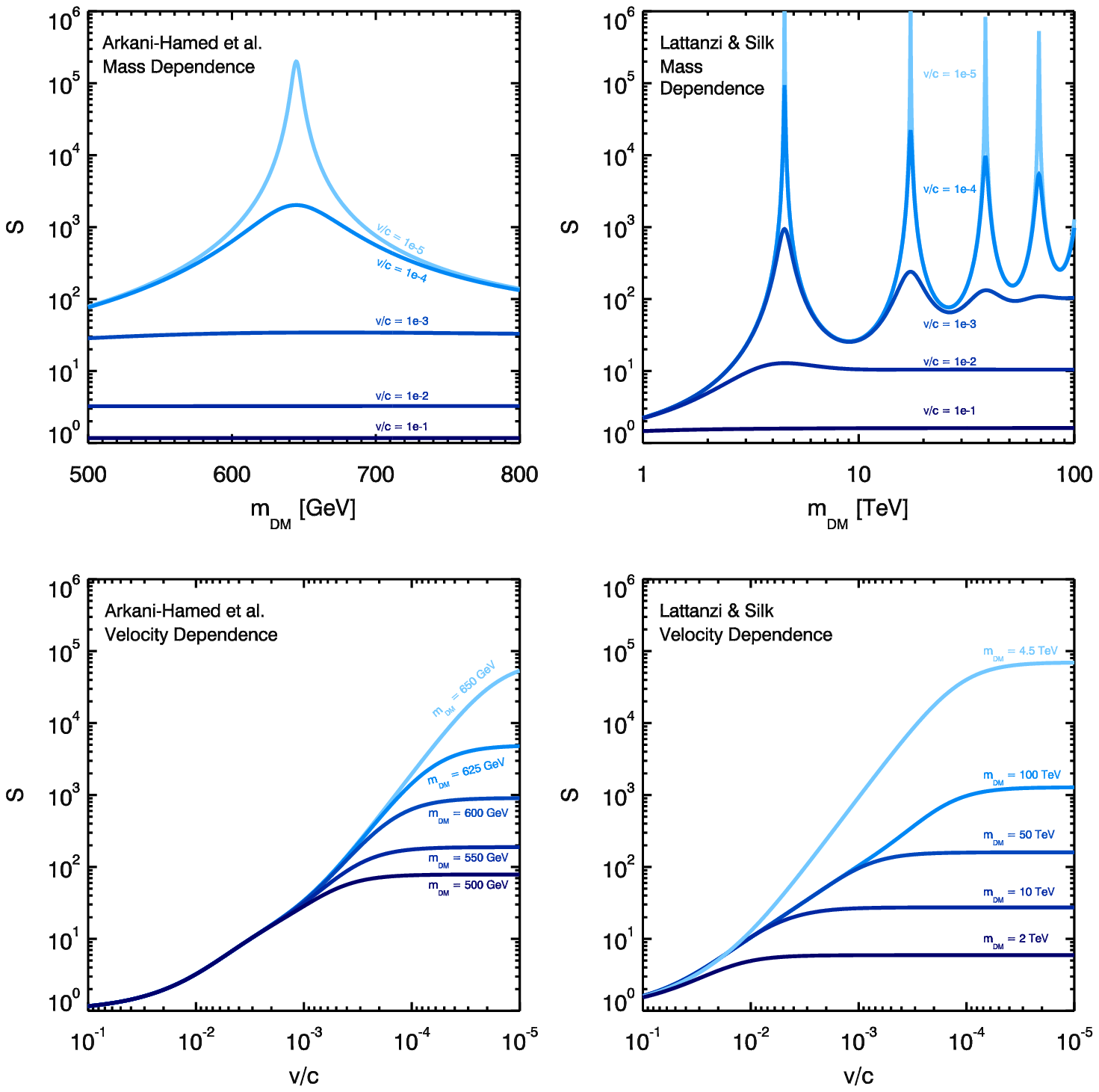}
\caption{\label{fig:sommerfeld}
Sommerfeld enhancement to the interaction cross section for annihilation using 
the parameters of the \citet{arkani-hamed2008a} model ({\em left column}, 
$\alpha\approx0.01$, $\mboson\sim1$~GeV) and \citet{lattanzi2008a} model 
({\em right column}, $\alpha\approx1/30$, $\mboson\approx90$~GeV).   
We show the cross section enhancement as a function of DM particle
mass $\mDM$ ({\em upper row}) or particle velocity 
({\em lower row}).  The right panels reproduce the results shown 
in Figures 2 and 3 of \citet{lattanzi2008a}.  
}
\end{center}
\end{figure*}

For the interaction of two particles with velocities
$\mathbf{v}_{1}$
and
$\mathbf{v}_{2}$, 
the collision rate is proportional to the integral
\begin{equation}
\label{equation:cross_section}
\langle\sigma v\rangle = \int \sigma(\vrel) \vrel f(\mathbf{v}_{\mathrm{rel}}) \dd^3 \mathbf{v}_{\mathrm{rel}},
\end{equation}
\noindent
where $\mathbf{v}_{\mathrm{rel}} \equiv \mathbf{v}_{1} - \mathbf{v}_{2}$.  
The relative velocity distribution $f(\vrel)$, which depends on the local
halo velocity dispersion $\sigmav(x)$, can be computed from the 
individual velocity distributions of particles by changing variables 
from $\mathbf{v}_{1}$ and $\mathbf{v}_{2}$ to $\mathbf{v}_{\mathrm{rel}}$ and the 
center-of-mass velocity $\mathbf{v}_{\mathrm{cm}} =\mathbf{v}_{1} + \mathbf{v}_{2}$ and 
integrating over $\mathbf{v}_{\mathrm{cm}}$ and the directional dependence of $\mathbf{v}_{\mathrm{rel}}$.

To illustrate the importance of the radial dependence of $\sigmav$, we 
approximate the velocity distribution function as an isotropic Maxwellian distribution with dispersion 
$\sigmav \ll \vesc \equiv \sqrt{-2\Phi}$.  In this case, the relative velocity 
distribution is simply 
\begin{equation}
\label{equation:fvrel}
f(\vrel) = \vrel^{2} \frac{1}{2\sqrt{\pi}\sigmav^{3}}\exp\left(-\frac{\vrel^{2}}{4\sigmav^{2}}\right).
\end{equation}
\noindent
We emphasize here that we utilize this form as a matter of convenience.  
Halo velocity distributions generally exhibit both velocity anisotropy and 
deviations from the Maxwellian form, but can be calculated directly from
the distribution function (see \S~4 or Ref.~\cite{binney1987a}).
For our purposes, the Maxwellian approximation should be conservative
since the kurtosis of the velocity distribution in the halo models we
adopt increases beyond the Gaussian value at radii $x\lesssim0.1$
\citep[see, e.g., ][]{magorrian1994a} and would result in a further
enhancement of the annihilation rate for the cross section velocity-dependence
we consider below.
We expect that the effect of velocity anisotropy will be to introduce 
geometrical factors of order unity into the evaluation of 
Eqn. \ref{equation:cross_section} by altering the form of the distribution
function.

\section{Sommerfeld Cross section Enhancement}
\label{section:sommerfeld}

As we argue above, if the particle interaction cross section has
some additional velocity dependence beyond the standard
$\langle \sigma v\rangle \sim$ constant, the kinematical structure
of the DM halo may alter the annihilation rate.  In what
follows, we consider the Sommerfeld enhancement of annihilation 
cross sections as a particular example of a velocity-dependent 
annihilation cross section of significant current interest.

The \citet{sommerfeld1931a} enhancement describes the increase in the
effective cross section owing to an attractive force between incident 
particles.  Let $\psi(r)$ denote the radial wavefunction of the 
equivalent one-body problem relative to the center-of-mass.  
The annihilation rate should be proportional to $\vert \psi(r) \vert^2$ 
in a region near the origin.  The Sommerfeld enhancement is an 
increase in the annihilation rate when the wavefunction $\psi(r)$ 
near the origin is significantly altered owing to an additional, 
relatively long-range interaction.  Let $\psi_0(r)$ and $\sigma_0$ be 
the radial wavefunction and annihilation cross section absent any new 
long-range force.  Upon introduction of the new force, the effective 
cross section is shifted to $\sigma = S \sigma_0$, where 
$S = \vert \psi(r=0)/\psi_0(r=0)\vert^2$ is the Sommerfeld factor.  
The Sommerfeld factor can be calculated from the attenuation of the 
solution $\psi(r)$ using the optical theorem as 
\cite{hisano_etal05,arkani-hamed2008a,lattanzi2008a} 
\begin{equation}
\label{equation:sommerfeld}
S = \frac{|\psi(\infty)|^{2}}{|\psi(0)|^{2}}.
\end{equation}
\noindent

If we assume s-wave ($\ell=0$) annihilation in the non-relativistic limit, 
then the scattering wavefunction is a solution to the Schr\"odinger equation 
in the form, 
\begin{equation}
\label{equation:schrodinger}
\frac{1}{\mDM} \frac{\dd^{2} \psi(r)}{\dd r^{2}} - V(r)\psi(r) = -\mDM \left(\frac{v}{c}\right)^2 \psi(r),
\end{equation}
\noindent
where $\mDM$ is the DM particle mass.  
If the attractive force has a finite range, Eq.~(\ref{equation:schrodinger})
can be solved numerically using the boundary condition that 
far from the origin the particle is a free wave, and
\begin{equation}
\psi(r\to\infty) \propto \exp\left( i \frac{\mDM v}{c} r\right).
\end{equation}
\noindent
Once the solution to $\psi(r)$ is integrated inward
to $r \to 0$, the velocity-dependent Sommerfeld enhancement 
can be evaluated from Eq.~(\ref{equation:sommerfeld}).

\begin{figure*}
\begin{center}
\includegraphics[height=9cm]{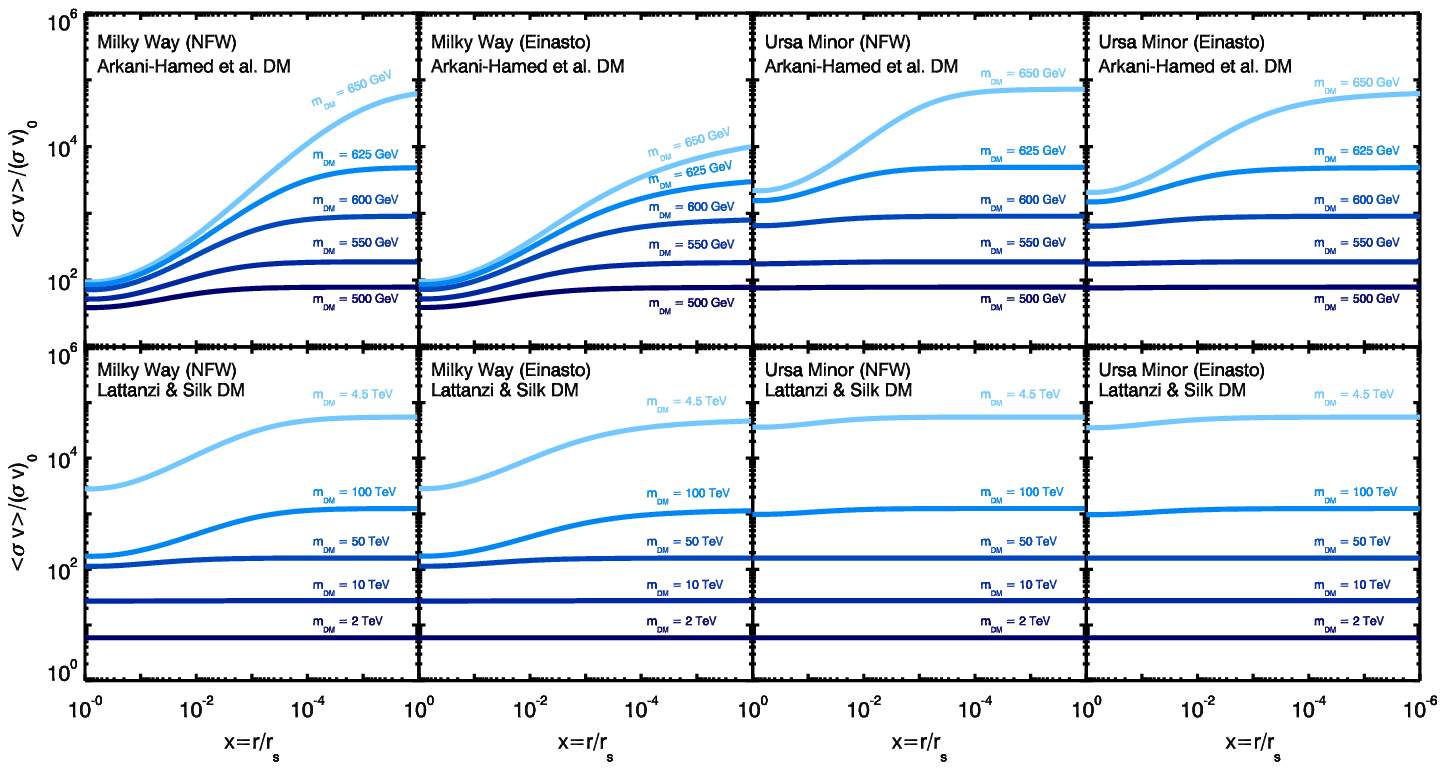}
\caption{\label{fig:cross_section}
Radial enhancement in the DM annihilation rate owing to a Sommerfeld-like 
velocity-dependent cross section and the radially-dependent halo velocity dispersion $\sigmav$.
Shown are the products of cross section and velocity averaged over the local velocity
distribution function and normalized to their values absent any Sommerfeld effect, 
$\langle\sigma v\rangle/(\sigma v)_0 = \int S(v) f(v) \dd v$.  
Position is given in units of halo scale radii, $x=r/\rs$.  
The annihilation enhancement is plotted for MW-like ({\em left panels}) and 
Ursa Minor-like ({\em right panels}) halos with NFW ({\em 1st and 3rd columns}) 
and Einasto ({\em 2nd and 4th columns}) halo profiles,  
using either the \citet{arkani-hamed2008a} or \citet{lattanzi2008a} DM parameters.\ 
}
\end{center}
\end{figure*}

For definiteness, we take the interaction to be given by an attractive 
Yukawa potential
\begin{equation}
\label{equation:yukawa}
V(r) = -\frac{\alpha}{r} \exp\left(-\mboson r\right)
\end{equation}
\noindent
where $\alpha>0$ is the coupling strength and
$\mboson$ is the mass of the boson that mediates the new force.  
This useful model for understanding the phenomenology of the 
Sommerfeld effect has been explored in Refs.~\cite{arkani-hamed2008a,lattanzi2008a}.  
$S$ can be large when $ v/c \ll \alpha$ and saturates to a maximum 
value for relative speeds 
\begin{equation}
v/c \ll \sqrt{\alpha \mboson/\mDM}.
\label{equation:saturation}
\end{equation}
Large enhancements require $\sqrt{\alpha \mboson/\mDM} \lesssim v/c \lesssim \alpha$.
In general, bound states exist when $\mDM/\mboson \sim n^2/\alpha$, 
where $n$ is an integer, and lead to large, resonant effective 
cross section enhancements.  

For our demonstration, we assume that the cross section can be written as 
$\sigma(v)v = \sigmavz S(v)$ where $\sigmavz$ is independent of 
velocity and the Sommerfeld factor $S(v)$ contains the entire velocity dependence.  
In the Yukawa model, the Sommerfeld factor additionally depends on the ratio.
$\mboson/\mDM$.  
As relevant examples, we will use the DM model parameters of 
\citet{arkani-hamed2008a} ($\alpha = 10^{-2}$, $\mboson = 1$~GeV, and $\mDM = 500-800$~GeV) 
and \citet{lattanzi2008a} ($\alpha = 1/30$, $\mboson = 90$~GeV, and $\mDM = 1-100$~TeV).  
The parameters of Ref.~\cite{arkani-hamed2008a} were 
chosen to match the mass scale of several hundred GeV selected by the ATIC data \cite{chang2008a} 
while also satisfying Eq.~(\ref{equation:saturation}) so that large boosts 
are attainable ($v/c \lesssim 10^{-3}$ in the local dark matter halo).  
The Ref.~\cite{lattanzi2008a} parameters 
are based on the same considerations for large $S$-factors obtained via an 
interaction mediated by Z boson exchange.  
Though we make specific choices including the form of the interaction and 
the masses of particles, this method can be applied to any model of particle
interactions.
The qualitative features we describe hold for Sommerfeld-enhanced annihilation 
in general.

Figure~\ref{fig:sommerfeld} shows the dependence of the Sommerfeld enhancement 
on the dark matter particle masses and relative velocities (the right panels 
of Fig.~\ref{fig:sommerfeld} reproduce the results in Figures 2 and 3 of 
Ref.~\cite{lattanzi2008a}).  Sommerfeld factors can clearly be quite large 
for reasonable parameter values.  The upper panels of Fig.~\ref{fig:sommerfeld} 
show examples of resonant capture at specific values of $\mDM$.  The lower 
panels show that the Sommerfeld factor becomes unimportant as $v/c \to 1$ 
and saturates according to Eq.~(\ref{equation:saturation}).  For 
$\alpha \gtrsim v/c \gtrsim \sqrt{\alpha \mboson/\mDM}$ the Sommerfeld 
factor exhibits a strong relative speed dependence.  $S(v) \propto v^{-1}$ 
for models away from resonance while $S(v) \propto v^{-2}$ near resonances 
\cite{arkani-hamed2008a,lattanzi2008a}.

\section{The Enhancement of Velocity-Dependent Dark Matter Annihilation in Galactic Halos}
\label{section:examples}

Figure~\ref{fig:cross_section} shows the enhancement of the annihilation 
cross section owing to the Sommerfeld effect as a function of position 
in dark matter halos.  We show the quantity 
$\langle \sigma(v) v \rangle/\sigmavz = \int S(v) f(v) \dd v$, 
which amounts to the Sommerfeld factor averaged over the 
relative velocity distribution functions for dark matter halo models 
at each position within the halo.  We show halo models representative of 
the MW and the Ursa Minor satellite for the dark matter parameters 
chosen by \citet{arkani-hamed2008a} and \citet{lattanzi2008a}.

The behavior of the annihilation enhancement 
$\langle \sigma v\rangle/\sigmavz$ 
depends on the halo kinematical structure and the details of the 
particle model.  If the DM particle mass is near resonance, 
the Sommerfeld enhancement increases faster than 
$S(v)\propto v^{-1}$ and the decline of the halo velocity dispersion
with radius leads to a large increase in the annihilation rate between 
the radii $x \sim 1$ and $x \sim 0$ {\em in addition} to the 
expected $\Gamma \propto \rho^2$ dependence.  
The relative increase is larger for the Milky Way 
($\langle \sigma v \rangle_{x\sim0}/\langle \sigma v\rangle_{x\sim1} \sim10^{3}$ 
for \citet{arkani-hamed2008a} DM with mass $\mDM\approx650$~GeV) 
than for Ursa Minor 
($\langle \sigma v \rangle_{x\sim0}/\langle \sigma v\rangle_{x\sim1} \sim30$) 
because the velocity dispersion of Ursa Minor near $x \sim 1$ 
is already approaching the velocity at which the Sommerfeld boost saturates.   
The faintest known Local Group dwarfs 
such as SEGUE 1 \cite{geha2008a} or Coma Berenices \cite{belokurov2007a}
with very low velocity dispersions 
($\sigmav\lesssim10$~km/s, see \citet{martinez2009a}) 
will have a radially-dependent cross section enhancement only
if the saturation velocity is very low ($v/c\lesssim10^{-5}$).
When $\mDM$ is such that the interactions are far from resonance, the 
effects are much less dramatic for the parameters we have chosen, but 
can be significantly larger in models with either smaller coupling strength $\alpha$ 
or larger $\mDM$, so that saturation occurs at significantly lower velocities 
as given by Eq.~(\ref{equation:saturation}).  

One aspect of Fig.~\ref{fig:cross_section} is worthy of explicit note in the context of 
recent proposals to explain the PAMELA/ATIC data with DM annihilations.  In 
the MW halo, the diffusion length of multi-GeV positrons is short ($\lesssim$~kpc) 
so that the relevant enhancement is the enhancement evaluated in a region local 
to the solar neighborhood, with $x \approx 0.35$.  Fig.~\ref{fig:cross_section} 
indicates that for a fixed enhancement factor in the Solar neighborhood, the 
enhancement factor toward the Galactic center can vary significantly depending 
upon the specifics of the model.  For these examples, the effective cross section at 
the Galactic center can be $\sim10-10^{3}$ times larger than its local
value.  

Similarly, the case of Ursa Minor as an example of a Local Group dwarf 
satellite is interesting.  
Should observable $\gamma$-rays be produced 
in this satellite, the 
surface brightness profile will result from the
position-dependent density {\it and} velocity distribution.
In models with Sommerfeld enhancements, the radial
gradient in the annihilation signal encodes information about the
mass ratio $\mboson/\mDM$.  
The scale radius of Ursa Minor subtends $\sim 2^{\circ}$ on the sky, 
and so this gradient may be a challenge to identify with an 
instrument such as FGST, with an angular resolution of 
$\sim 10$~arcminutes, but may be accessible to atmospheric 
Cerenkov detectors such as VERITAS, HESS, MAGIC, and CANGAROO 
with angular resolutions of the order of arcminutes.

\section{Discussion and Conclusions}
\label{section:discussion}

We have studied the influence of the kinematical structure 
of dark matter halos on dark matter annihilations in models 
with velocity-dependent annihilation cross sections.  
The methods we have used are general, but we have considered 
specific examples of the Sommerfeld enhancement to DM annihilations 
that have been proposed in order to explain recent measurements of 
high energy positron fluxes by PAMELA and ATIC.  As our primary 
aim has been to illustrate the general importance of halo kinematics 
for velocity-dependent annihilations, we have cast our answers 
in terms of relative annihilation cross sections without specifying 
either a normalization of the cross section at high relative 
velocity [$\sigmavz$] or particular yields of outgoing leptons 
and photons from such annihilations.
Our results demonstrate that the kinematical structure of 
DM halos can influence the annihilation rate in important ways.  
In particular, the velocity distributions of dark matter particles 
may be a strong function of position so that velocity-dependent 
annihilation cross sections may give rise to annihilation rates 
that vary with halo position in addition to the dependence upon local number 
density $\Gamma \propto \rho^2(r)$.  
These results could easily be incorporated into cosmological
N-body simulation estimates of dark matter annihilation rates 
\citep[e.g.,][]{kuhlen2008a,springel2008a}.

Halo velocity structure may be important for annihilations within the MW.  
For the specific cases of Sommerfeld enhancements 
with the \citet{arkani-hamed2008a} and \citet{lattanzi2008a} DM parameter choices, 
halo kinematical structure leads to an effective DM annihilation cross section 
that rises by a factor of $\sim 10-10^3$ between the solar neighborhood and 
the MW halo center.  The increase in the annihilation
rate is most prevalent when the interaction occurs near resonance with 
$\mDM \sim n^2 \mboson/\alpha$.  In any case, the enhancement 
tends to a constant at small radii owing to the saturation of the 
Sommerfeld effect when condition Eq.~(\ref{equation:saturation}) is met. 
This radial dependence can lead to relatively large boosts in 
annihilation products from the Galactic center relative to the local 
boost with or without taking additional boosts from 
halo substructure into consideration.  

The examples that we have presented are simple, 
but we have studied more complete models of the Milky Way 
including the influence of the baryonic components of the Milky Way 
galaxy according to the parameterizations of 
\citet{klypin2002a} and self-consistent models of \citet{widrow2005a}.  
In all cases, our qualitative result holds.  
The kinematical structure of the DM halo remains important in the 
sense that the Sommerfeld boost in the solar neighborhood can be 
significantly different from the boost elsewhere in the halo.  What 
differs in these models are the detailed parametrics.  For a specific 
set of dark matter interaction parameters, including the potential 
of the baryonic component of the Galaxy decreases the local Sommerfeld 
boost but increases the gradient of $\langle \sigma v \rangle/\sigmavz$.  
Including a large black hole at the center of the Galactic potential 
leads to a cut-off in the enhancement level at roughly the radius 
of influence of the black hole, which is of order $r \sim 10^{-3}-10^{-4} \rs$ 
and depends in detail upon the structural parameters one assumes for the 
galaxy and the halo.  At this early stage, a detailed exploration of the 
parameter space seems of limited value, though we speculate that it may 
be more fruitful to seek such tests of velocity dependent cross sections 
in dwarf galaxies where there are fewer complications.

The dwarf satellites of the Milky Way may be additional sources of 
observable annihilation radiation that can shed light on the effective 
cross section for DM annihilation \cite{arkani-hamed2008a}.  We have 
considered the case of the dwarf satellite Ursa Minor because it 
subtends a large solid angle on the sky and would be one of the most 
luminous of any known MW satellite in DM annihilation products \cite{strigari2008a}.  
Our results indicate that in systems like Ursa Minor
the kinematical structure of the halo may 
reveal a signature of Sommerfeld-enhanced annihilation.  
In 
particular, models with large Sommerfeld enhancements to the dark 
matter annihilation cross section may result in a surface brightness 
profile that varies more rapidly than the density-squared weighting 
would imply over the inner few arcminutes of the galaxy.  The scale 
over which the variation is significant is small compared to the 
angular resolution of FGST ($\sim 10$~arcmin.), but comparable to 
that of atmospheric Cerenkov detectors such as HESS, MAGIC, 
VERITAS, and CANGAROO.

Of course, the absolute cross section for annihilation is not known, 
so models for which halo structure will be important 
are cases where the Sommerfeld factor is not yet in the saturation 
regime for a large fraction of the halo.  
In the case of the MW, 
a large radial gradient in the annihilation rate requires that saturation 
should occur at speeds less than 
the local DM velocity dispersion of $v/c \sim 10^{-3}$.  For dwarf halos 
such as Ursa Minor the requirement is $v/c \lesssim 10^{-4}$, while
smaller dwarfs like SEGUE 1 or Coma Berenices require saturation at $v/c \lesssim 10^{-5}$.  
For these conditions to be satisfied,
either the coupling constant $\alpha$ or 
the mass ratio $\mboson/\mDM$ must be tuned to values many orders magnitude 
below unity.  The Sommerfeld enhancement saturation explains why the spatial dependence of the effective 
cross section is mild in Ursa Minor for the \citet{lattanzi2008a} particle model shown
in Fig.~\ref{fig:cross_section}.  
Likewise, the first collapsing halos with masses for less than 
the halos of contemporary dwarf galaxies will lie well within the 
saturation regime (if $M \sim M_{\oplus}$, then $v/c \sim 10^{-8}$), 
so that constraints from $\gamma$-ray backgrounds owing to annihilation 
in these objects remain relevant \cite{profumo2006a,kamionkowski2008a}.

The position dependence of the annihilation cross section may be 
a testable prediction of Sommerfeld-enhanced DM annihilation models 
of the PAMELA/ATIC data in some regions of parameter space.  
For the Milky 
Way, the decreasing width of the velocity distribution can result in
far greater fluxes of annihilation products from the 
Galactic Center than would otherwise be expected.  In the 
case of Ursa Minor, it might be possible to detect the signature of 
an effectively spatially-dependent annihilation cross section that 
varies on scales of $\sim $~arcminutes.  
If the PAMELA/ATIC data have an astrophysical explanation 
\cite{busching_etal08,profumo2008a}, the current motivation 
for Sommerfeld boosts would be diminished, 
but an enhancement owing to a velocity-dependent cross section for interactions 
among dark matter particles may still be possible.  In such cases, the Sommerfeld 
boost may manifest solely as modified spatial profiles of 
$\gamma$-rays from halo particle annihilations.  Either way, it is 
an exciting time in the quest to identify the dark matter and the 
flood of data expected over the next few years will only add to this 
excitement.

\begin{acknowledgments}

We are grateful to Babu Bhatt, Dan Hooper, Marc Kamionkowski,
Savvas Koushiappas, Louis Strigari,
and Amol Upadhye for 
useful discussions.  This work was motivated in part by the Particle Cosmology
Stimulus Package meeting between the University of Chicago and 
Fermilab.  BER gratefully acknowledges support from a
Spitzer Fellowship through a NASA grant administrated by the Spitzer
Science Center, and is partially supported by
the Kavli Institute for Cosmological
Physics at the University of Chicago. 
ARZ is supported by the University of Pittsburgh, 
by the National Science Foundation through grant AST 0806367, and by 
the Department of Energy.

\end{acknowledgments}


\end{document}